\documentclass[RNAAS]{aastex62}
\usepackage{comment}
\usepackage{graphicx}
\usepackage{natbib}
\usepackage{amsmath,amssymb}
\usepackage{parskip}
\usepackage{tikz}
\usetikzlibrary{intersections} 

\graphicspath{{./}{figures/}}

\begin{document}

\title{An Alternative Derivation of the Analytic Expression of Transmission Spectra}

\correspondingauthor{Andr\'es Jord\'an}
\email{ajordan@astro.puc.cl}

\author[0000-0002-5389-3944]{Andr\'es Jord\'an}
\affiliation{Instituto de Astrof\'isica, Pontificia Universidad Cat\'olica de Chile, Av.\ Vicu\~na Mackenna 4860, Macul, Santiago, Chile}
\affiliation{Millennium Institute of Astrophysics, Chile}

\author[0000-0001-9513-1449]{N\'estor Espinoza}
\altaffiliation{Bernoulli fellow; Gruber fellow}
\affiliation{Max-Planck-Institut f\"ur Astronomie, K\"onigstuhl 17, Heidelberg 69117, Germany }


\section{} 

Under some assumptions, an analytic expression for the transmission spectrum can be obtained \citep{deWit:2013,betre:2017}, which can form the basis of atmospheric retrievals and allows insight on the degeneracies involved \citep{heng:2017}.
In this Research Note we present an alternative derivation for the analytic expression of a transmission spectrum first derived in \citet{deWit:2013}. 

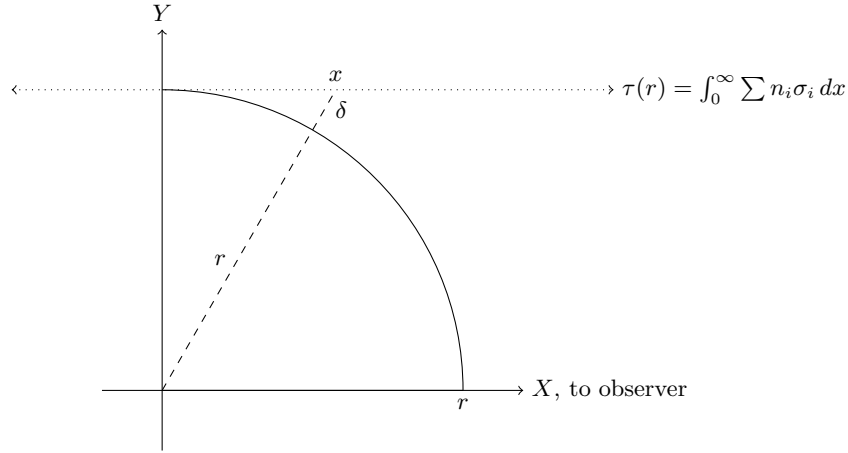
\begin{figure}[h!]
\centering
\begin{tikzpicture}[scale=4]
\draw[->] (-0.2,0) -- (1.2,0) node[right] {$X$, to observer};
\draw[->] (0,-0.2) -- (0,1.2) node[above] {$Y$};
\draw (0,0) -- (0:1);
\draw (90:1) arc (90:0:1) node[below] {$r$};
\draw[<->,dotted] (-0.5,1) -- (1.5,1) node[right] {$\tau(r) = \int_0^\infty \sum n_i\sigma_i \, dx$};
\draw[dashed] (0,0) -- node[left=1pt] {$r$} (60:1);
\path [name path=diagonal line] (0,0) -- (60:1.2);
\path [name path=slanted] (0,1) -- (1,1);
\draw [name intersections={of=diagonal line and slanted, by=x}, dashed] (60:1) -- node[right=1pt] {$\delta$} (x) node[above] {$x$};
\end{tikzpicture}
\caption{Illustration of the geometry.}
\label{fig:geom}
\end{figure}

The equation defining the transit radius as a function of wavelength $\lambda$ is 

\begin{equation}
\pi R_p(\lambda)^2 = \int_0^\infty 2\pi r[1-\exp(-\tau(r,\lambda))]\,dr
\label{eq:Rp}
\end{equation}

\noindent where $\tau(r,\lambda)$ is the slant optical depth through the atmosphere 
which reaches minimum distance $r$ to the center, i.e.

\begin{equation}
\tau(r, T, P,\lambda) = 2 \int_0^\infty \sum_{i=0}^N n_i(r+\delta_{r,x})\sigma_i(r+\delta_{r,x},T,P, \lambda) \,dx
\end{equation}

\noindent where we consider $N$ species with number densities $n_i(r)$ and cross sections $\sigma_i(r,T,P, \lambda)$ (see Figure~\ref{fig:geom} for the geometry and definition of $\delta_{r,x}\equiv\delta(r,x)$). It is customary to assume at this point an isothermal atmosphere and that  species behave as ideal gases. Hydrostatic equilibrium implies then $P(z) = P_r\exp(-z/H)$, where $P_r$ is the  pressure  at some arbitrary reference radius from where the altitude $z$ is measured, and $H$, the atmospheric scale-height, is  given by $H = k_BT/\mu g$, where $\mu$ is the mean molecular weight, $g$ the surface gravity, $T$ the temperature and $k_B$ Boltzmann's constant. Because of the isothermal assumption we also have that $n \propto \exp(-z/H)$.
The atmosphere is assumed to be thin enough that we can assume $g \approx$ constant. The same assumption of a thin atmosphere leads to the approximation $x^2 + r^2 = (r+\delta_{r,x})^2 \approx r^2 + 2r\delta_{r,x}$, i.e.\ $\delta_{r,x} = x^2/2r$, and we can then write $n_i(r+\delta_{r,x}) \approx n_i(r)\exp(-x^2/(2rH))$ \citep{fortney:2005}. 
Assuming also that $\sigma_i$ does not depend on position, either implicitly through $T,P$ or explicitly, we can easily carry out the integral over $x$, arriving at\footnote{We are considering only absorption that is proportional to the number density, e.g.\ Rayleigh scattering. We could also incorporate processes with absorption $\propto n^q$, such as collisionally induced absorption ($q=2$), in which case $\tau \propto n_0^q \sqrt{2\pi r H/q}$, see \S2.2 \citet{betre:2017}.}

\begin{equation}
\tau(r,\lambda) = \sum_{i=0}^N \sigma_i(\lambda) n_i(r)\sqrt{2\pi r H}.
\label{eq:tau}
\end{equation}

Note that a ``surface" radius $R_0$ does not appear in this expression. $R_0$ is a variable introduced in all extant derivations at this stage as a means to decompose $r$ as $R_0 + z$, and assume that $\sqrt{2\pi r H}$ is approximately constant and equal to $\sqrt{2\pi R_0 H}$. 
The next step is to use the relation $\tau=\sum_{i=0}^N \sigma_i(\lambda) n_i(r)\sqrt{2\pi R_0 H}$ to change variables in the integral from $r$ to $\tau$. 
If we insist in maintaining $r$ in the square root we cannot follow this path, but  making such change of variables is actually not needed to derive the result, so the introduction of $R_0$ is not strictly necessary at the stage done in existing treatments.

Let's go back to equation~\ref{eq:Rp}, and use the expression for $\tau$ from equation~\ref{eq:tau}. We denote $\sqrt{2\pi r H} \equiv \alpha r^{1/2}$ and $\sum_{i=0}^N \sigma_i(\lambda) n_i(r) = n(r)\sum_{i=0}^N \chi_i\sigma_i \equiv n(r) \beta(\lambda)$, where $\chi_i$ are volume mixing ratios and $n$ is the total number density. 
Given our assumptions, we can write $n(r) = \zeta \exp(-r/H)$, where $\zeta$ is a constant. Clearly, this expression is valid only within the atmosphere and it would be an unwarranted extrapolation using this expression of $n(r)$ throughout the whole integral in equation~\ref{eq:Rp}. It is at this point that specifying a ``surface" becomes necessary if one is to carry out the integral in equation~\ref{eq:Rp}: the integral is no longer from $r=0$ but from some value $R_0$ from where the isothermal atmosphere expression for $n(r)$ is a good approximation. 
In what follows $R_0$ satisfies the condition of being a radius below which the planet is fully opaque\footnote{That $R_0$ must satisfy this is sometimes lost on the literature. For example, \citet{heng:2017} write ``In the current derivation, $\tau_0$ 
[$n_0 \sqrt{2\pi H R_0}$] is simply associated with a reference pressure corresponding to an atmospheric layer that may -- or may not -- be chosen to be optically thick". But this is not correct, their derivation, as well as ours, has to assume that the surface radius $R_0$ is optically thick. This assumption is made implicitly when $\pi R_p^2$ is written as $\pi R_0^2 + A$, where $A$ is the integral in equation~\ref{eq:Rp_split} above. If $R_0$ is not chosen to be at an optically thick region (i.e., a region where $\tau \to \infty$), then it is not possible to write $2\pi \int_0^{R_0} r[1-\exp(-\tau)] \,dr = \pi R_0^2$. } and above which the isothermal expression for pressure and density is valid. 
We have then that $R_p(\lambda)$ is given by

\begin{equation}
\pi R_p(\lambda)^2 = \pi R_0^2 +  \int_{R_0}^\infty 2\pi r[1-\exp(-\tau(r,\lambda))]\,dr
\label{eq:Rp_split}
\end{equation}

Above $R_0$, by construction we can assume that $n(r) = \zeta \exp(-r/H)$. A convenient way to define $\zeta$ is $\zeta= n_0 \exp(R_0/H)$, where $n_0$ is the total number density at $R_0$. Using these definitions,

\begin{align}
\pi (R_p(\lambda)^2-R_0^2) &= \int_{R_0}^\infty 2\pi r[1-\exp(-\sum_{i=0}^N \sigma_i(\lambda) n_i(r)\sqrt{2\pi r H})]\,dr \\
	&= 2\pi \int_{R_0}^\infty r[1-\exp(- \beta(\lambda) \zeta \alpha \exp(-r/H)r^{1/2})]\,dr 
\end{align}

Defining $\xi_\lambda = \beta(\lambda) \zeta \alpha$, 

\begin{align}
\pi (R_p(\lambda)^2-R_0^2) &= 2\pi \int_{R_0}^\infty r\left[\sum_{k=1}^\infty (-1)^{k+1}\xi_\lambda^k \exp(-kr/H)r^{k/2}/k!\right] \,dr \\
    &= 2\pi \sum_{k=1}^\infty (-1)^{k+1}\frac{\xi_\lambda^k}{k!} \int_{R_0}^\infty \exp(-kr/H)r^{k/2+1} dr\\
    &= 2\pi  \sum_{k=1}^\infty (-1)^{k+1}\frac{\xi_\lambda^k}{k!} \left(\frac{H}{k}\right)^{k/2+2} \int_{kR_0/H}^\infty \exp(-u)u^{k/2+2 -1} du \\
    &= 2\pi  \sum_{k=1}^\infty (-1)^{k+1} \frac{\xi_\lambda^k}{k!} \left(\frac{H}{k}\right)^{k/2+2} \Gamma(k/2 + 2,kR_0/H) 
\end{align}

\noindent where $\Gamma(\alpha,x) = \int_x^\infty e^{-t}t^{\alpha-1}\,dt$  is the incomplete Gamma function. Now, $kR_0/H \gg 1$ so we can use the following asymptotic expansion of the incomplete Gamma function \citep[][8.357]{gradshteyn}

\begin{equation}
\Gamma(\alpha,x) = x^{\alpha-1}\exp(-x)\left[\sum_{m=0}^{M-1}\frac{(-1)^m\Gamma(1-\alpha+m)}{x^m \Gamma(1-\alpha)} + \mathcal{O}(|x|^{-M})\right]
\end{equation}

\noindent valid for large values of $|x|$. We have then that

\begin{align}
\pi (R_p(\lambda)^2-R_0^2) &= 2\pi  \sum_{k=1}^\infty (-1)^{k+1} \frac{\xi_\lambda^k}{k!} \left(\frac{H}{k}\right)^{k/2+2} \left(\frac{kR_0}{H}\right)^{k/2+1} \exp(-kR_0/H) \times\\
         & \left[ \sum_{m=0}^{M-1}\frac{(-1)^m\Gamma(m-k/2-1)}{(kR_0/H)^m \Gamma(-1-k/2)} + \mathcal{O}(|kR_0/H|^{-M})\right]. 
 \label{eq:full_expansion}
\end{align}

In order to arrive to a closed analytical expression, we keep only the first order $M=1$, which gives

\begin{align}
\pi (R_p(\lambda)^2-R_0^2) &= 2\pi  \sum_{k=1}^\infty (-1)^{k+1} \frac{\xi_\lambda^k}{k!} \left(\frac{H}{k}\right)^{k/2+2} \left(\frac{kR_0}{H}\right)^{k/2+1} \exp(-kR_0/H) \times \\
         & \left[ 1 + \mathcal{O}(|kR_0/H|^{-1})\right]. \nonumber
\end{align}

We have then 

\begin{align}
\pi (R_p(\lambda)^2-R_0^2) &= 2\pi  \sum_{k=1}^\infty (-1)^{k+1} \frac{\xi_\lambda^k}{k!} \left(\frac{H}{k}\right)^{k/2+2} \left(\frac{kR_0}{H}\right)^{k/2+1} \exp(-kR_0/H)\\
    &= 2\pi H R_0\sum_{k=1}^\infty (-1)^{k+1} \frac{\xi_\lambda^k}{k!} R_0^{k/2} \left(\frac{1}{k}\right) \exp(-kR_0/H).
\end{align}

We now make use of the following series representation of the exponential-integral function $Ei(x)$ \citep[][8.214]{gradshteyn}

\begin{equation}
Ei(x) = \gamma + \ln(-x) + \sum_{k=1}^\infty \frac{x^k}{k\cdot k!} \,\,\,\, \mathrm{for}\,\,\,x<0
\end{equation}

\noindent where $\gamma$ is Euler-Mascheroni's constant, expand $\xi_\lambda = \beta(\lambda) \zeta \alpha = n_0 \exp(R_0/H) \sqrt{2\pi H} \beta(\lambda)$, and use the fact that $E_1(x) = -Ei(-x)$ for $x>0$, where $E_1$ is the exponential integral \citep[][Appendix I]{chandrasekhar}, to get

\begin{align}
\pi (R_p(\lambda)^2-R_0^2) &= -2\pi H R_0 \sum_{k=1}^\infty \frac{(-n_0 \exp(R_0/H) \sqrt{2\pi H} \beta(\lambda)\exp(-R_0/H) \sqrt{R_0})^k}{k \cdot k!}\\
        &= -2\pi H R_0 \sum_{k=1}^\infty \frac{(-n_0 \beta(\lambda) \sqrt{2\pi H R_0})^k}{k \cdot k!}\\
        &= 2\pi H R_0 \left[\gamma + \ln(n_0 \beta(\lambda) \sqrt{2\pi H R_0}) - Ei(-n_0 \beta(\lambda) \sqrt{2\pi H R_0})\right]\\        
        &= 2\pi H R_0 \left[\gamma + \ln(n_0 \beta(\lambda) \sqrt{2\pi H R_0}) + E_1(n_0 \beta(\lambda) \sqrt{2\pi H R_0})\right]
\label{eq:R_p_canon}
\end{align}

This is our final expression, which is equivalent to the expressions present in the literature.

\bibliographystyle{aasjournal}
\bibliography{transpec}

\end{document}